\documentclass[aip,preprint]{revtex4-1}
\draft 
\usepackage{graphicx}
\usepackage{bm}
\usepackage{color}

\begin{document}


\title{Charge-carrier mobility in hydrogen-terminated diamond field-effect transistors} 



\author{Yosuke Sasama}
\affiliation{International Center for Materials Nanoarchitectonics, National Institute for Materials Science, Tsukuba 305-0044, Japan}
\affiliation{Graduate School of Pure and Applied Sciences, University of Tsukuba, Tsukuba, 305-8571, Japan}

\author{Taisuke Kageura}
\affiliation{International Center for Materials Nanoarchitectonics, National Institute for Materials Science, Tsukuba 305-0044, Japan}

\author{Katsuyoshi Komatsu}
\affiliation{International Center for Materials Nanoarchitectonics, National Institute for Materials Science, Tsukuba 305-0044, Japan}

\author{Satoshi Moriyama}
\affiliation{International Center for Materials Nanoarchitectonics, National Institute for Materials Science, Tsukuba 305-0044, Japan}

\author{Jun-ichi Inoue}
\affiliation{International Center for Materials Nanoarchitectonics, National Institute for Materials Science, Tsukuba 305-0044, Japan}

\author{Masataka Imura}
\affiliation{Research Center for Functional Materials, National Institute for Materials Science, Tsukuba 305-0044, Japan}

\author{Kenji Watanabe}
\affiliation{Research Center for Functional Materials, National Institute for Materials Science, Tsukuba 305-0044, Japan}

\author{Takashi Taniguchi}
\affiliation{Research Center for Functional Materials, National Institute for Materials Science, Tsukuba 305-0044, Japan}

\author{\\ Takashi Uchihashi}
\affiliation{International Center for Materials Nanoarchitectonics, National Institute for Materials Science, Tsukuba 305-0044, Japan}

\author{Yamaguchi Takahide}
\affiliation{International Center for Materials Nanoarchitectonics, National Institute for Materials Science, Tsukuba 305-0044, Japan}
\affiliation{Graduate School of Pure and Applied Sciences, University of Tsukuba, Tsukuba, 305-8571, Japan}


\date{\today}

\begin{abstract}

Diamond field-effect transistors (FETs) have potential applications in power electronics and high-output high-frequency amplifications. In such applications, high charge-carrier mobility is desirable for a reduced loss and high-speed operation. We recently fabricated diamond FETs with a hexagonal-boron-nitride gate dielectric and observed a high mobility above 300 cm$^{2}$V$^{-1}$s$^{-1}$.
In this study, we identify the scattering mechanism that limits the mobility of our FETs through theoretical calculations. 
Our calculations reveal that the dominant carrier scattering is caused by surface charged impurities with a density of $\approx$1$\times10^{12}$ cm$^{-2}$, and suggest that an increase in mobility over 1000 cm$^{2}$V$^{-1}$s$^{-1}$ is possible by reducing these impurities.

\end{abstract}


\maketitle 


\section{introduction}
Diamond has attracted much attention as a next-generation semiconducting material because of its excellent properties, including a wide-band gap, high thermal conductivity, high breakdown electric field, and high mobility\cite{Wor08}.
These properties enable field-effect transistors (FETs) to operate at high temperature with low-energy loss, to resist high voltage, and to be reduced in size.
Previously, FETs operating\cite{kaw14} at 400$^\circ$C and at a breakdown voltage\cite{kit17} above 2000 V were reported.
The diamond FETs in the previous studies were often fabricated using a hydrogen-terminated diamond surface, which exhibits $p$-type surface conductivity.

$p$-type surface conductivity appears on hydrogen-terminated diamond after its exposure to air even if the diamond is non-doped.
The surface conductivity can be explained by the transfer doping model\cite{Mai00}.
The valence band maximum of hydrogen-terminated diamond is higher than the lowest unoccupied states of impurities adsorbing to the diamond surface due to air exposure.
The electrons in the valence bands are therefore transferred to the impurities, and holes are induced at the diamond surface, resulting in surface conductivity. 
It is known that transfer doping is also caused by exposure to NO$_{2}$ gas\cite{Kas17} and by deposition of a solid insulator with a high electron affinity, such as V$_{2}$O$_{5}$\cite{Ver18}.

The mobility corresponding to the surface conductivity of hydrogen-terminated diamond is typically 10-100 cm$^{2}$V$^{-1}$s$^{-1}$ at room temperature. This value is more than one order of magnitude lower than the intrinsic mobility of bulk diamond ($\approx$4000 cm$^{2}$V$^{-1}$s$^{-1}$)\cite{Isb02}.
Recently, Li {\it et al.} calculated the mobility of the surface conductivity of hydrogen-terminated diamond as a function of temperature and carrier density. They pointed out that the major scattering sources were surface impurities\cite{Li18}.

The mobilities of hydrogen-terminated diamond FETs are almost the same as or less than that of the surface conductivity of hydrogen-terminated diamond. This suggests that charged impurities also exist in the gate insulator.
In fact, it was suggested that Al$_{2}$O$_{3}$ film deposited by the atomic layer deposition (ALD) method contains oxygen point defects and aluminum vacancies, and electrons trapped in these defects are balanced with hole carriers in diamond\cite{Kaw17}.
Other gate insulators, such as SiO$_{2}$ and CaF$_{2}$, have been formed by vacuum evaporation or sputter deposition, but these films are amorphous and may also contain charged impurities and traps.

Recently, we fabricated diamond FETs using monocrystalline hexagonal boron nitride (h-BN) as a gate insulator and observed a high mobility above 300 cm$^{2}$V$^{-1}$s$^{-1}$.\cite{Sas18} The high mobility is presumably due to the low density of charged impurities in h-BN.
In this study, we calculated the room-temperature mobility of diamond FETs as a function of carrier density to reveal the factor limiting the mobility of these FETs.

\section{modeling}
The equations we used for calculating the scattering rate were basically the same as those used in the paper by Li {\it et al.}\cite{Li18}
There are, however, two major differences between our calculation and theirs.
First, Li {\it et al.} assumed that the density of surface charged impurities was the same as the carrier density.
This is reasonable for the surface conductivity of hydrogen-terminated diamond because the negative charges of the surface impurities are balanced with the positive charges of the holes in diamond according to the transfer doping model.
In diamond FETs, however, the density of surface charged impurities is independent of the carrier density because the carrier density can be controlled by the gate voltage.
We therefore treated the density of the surface charged impurities as a constant.
This treatment causes a qualitative difference between the carrier density dependence of the mobility in our calculation and that in the paper of Li {\it et al.}

\begin{figure}
 \includegraphics[width=6.3truecm]{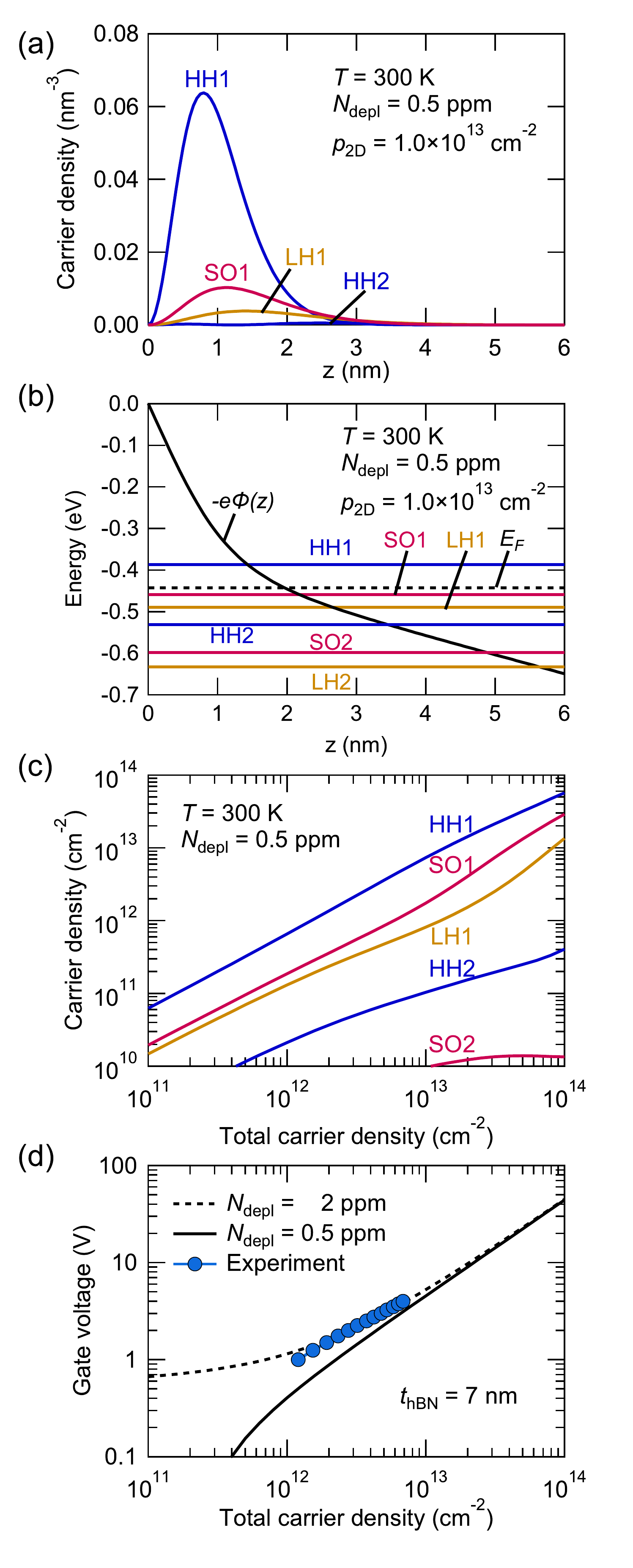}
 \caption{(a) Calculated density profile of the lowest subbands of heavy- (HH), light- (LH), and split-off (SO) holes and the second subband of heavy holes (HH2) for the total carrier density of 1.0$\times$10$^{13}$ cm$^{-2}$ at $T$ = 300 K. $z$ is the depth from the diamond surface. $N_{\rm depl}$ is assumed to be 0.5 ppm. (b) Calculated potential energy, Fermi level, and maximum of the first and second subbands of heavy, light, and split-off holes as a function of $z$. (c) Calculated population of heavy holes, light holes, and split-off holes as a function of total carrier density at $T$ = 300 K. (d) Calculated gate voltage vs. total carrier density for $t_{\rm hBN}=7$ nm thick gate dielectric. $N_{\rm D}=0.5$ ppm and $N_{\rm A}$=5 ppb for the solid line and $N_{\rm D}=2$ ppm and $N_{\rm A}=5$ ppb for the dashed line. $n_{\rm imp}=1.5\times10^{12}$ cm$^{-2}$ for both curves. Experimental results of our h-BN/diamond FET (Fig. 2(c) of Ref. 10) are also shown.}
 \label{fig:SP}
\end{figure}

The second point of difference is in the way of treating the three valence bands.
Li {\it et al.} used a single equivalent isotropic band model.
That is, they assumed a single effective band with a density-of-state mass $m_{\rm d}^{*}=((m^{\rm LH})^{3/2} + (m^{\rm HH})^{3/2} + (m^{\rm SO})^{3/2}))^{2/3}$ and conduction mass $m_{\rm c}^{*}=((m^{\rm LH})^{3/2} + (m^{\rm HH})^{3/2} + (m^{\rm SO})^{3/2}))/((m^{\rm LH})^{1/2} + (m^{\rm HH})^{1/2} + (m^{\rm SO})^{1/2}))$.
(They used the heavy hole (HH) mass $m^{\rm HH}=0.588m_{0}$, light hole (LH) mass $m^{\rm LH}=0.303m_{0}$, and split-off (SO) hole mass $m^{\rm SO}=0.394m_{0}$, with $m_{0}$ being the rest mass.)
In contrast, we distinguished the three bands and performed the mobility calculations for the three bands separately. The distribution of the carrier densities in the HH, LH, and SO bands was determined by solving the Schr\"{o}dinger and Poisson equations self-consistently,\cite{Ham17,Neb04,Edm10}
\begin{eqnarray}
 \left[-\frac{\hbar^{2}}{2m_{z}^{i}}\frac{d^{2}}{dz^{2}} + e\phi(z) (+\Delta^{\rm SO}) - E_{n}^{i}\right]\Psi_{n}^{i}(z) = 0,
 \label{eq:sh} \\
 \frac{d^{2}\phi(z)}{dz^{2}} = -\frac{1}{\epsilon_{0}\epsilon_{s}}\left[eN_{\rm depl} + e\sum_{i,n}p_{n}^{i}\left|\Psi_{n}^{i}(z)\right|^{2}\right],
 \label{eq:po} \\
 p_{n}^{i} = \frac{m_{//}^{i}k_{B}T}{\pi\hbar^{2}}\ln\left[1+\exp\left(\frac{E_{F}-E_{n}^{i}}{k_{B}T}\right)\right],
 \label{eq:nni} \\
 \sum_{i,n}p_{n}^{i} = p_{\rm 2D}.
 \label{eq:ni}
\end{eqnarray}
Here, $e$ is the elementary charge; $\epsilon _{0}$ is the vacuum permittivity; $\epsilon _{s}$ is the static dielectric constant of diamond; $e\phi(z)$ is the potential energy; $N_{\rm depl}=N_{D}-N_{A}$, where $N_{D}$ and $N_{A}$ are the ionized donor and acceptor concentrations in the diamond substrate; $E_{n}^{i}$ is the maximum energy of the $n$th sub-band; $\Psi_{n}^{i}(z)$ is the wave function corresponding to $E_{n}^{i}$; $p_{n}^{i}$ is the sheet hole density of the $n$th sub-band; $p_{\rm 2D}$ is the total sheet carrier density; $E_{F}$ is the Fermi level; $k_{B}$ is the Boltzmann constant; and $T$ is the absolute temperature.
We did not consider the band mixing of the HH, LH, and SO subbands.
Examples of $\Psi_{n}^{i}(z)$, $e\phi(z)$, $E_{n}^{i}$, and $E_{F}$, which are the solutions for $p_{\rm 2D}=1\times10^{13}$ cm$^{-2}$, are shown in Fig. \ref{fig:SP}(a) and \ref{fig:SP}(b).
The carrier densities ($p^{\rm HH}$, $p^{\rm LH}$, and $p^{\rm SO}$) for HH, LH, and SO holes were obtained by summing the densities over the seven lowest subbands ($p^{i} = \sum_{n=1}^{n_{\rm max}}p_{n}^{i}$, $n_{\rm max}=7$).
The sum of the first HH, LH, and SO subbands reaches $>96\%$ of the total carrier density in the range of total carrier density between 1$\times$10$^{11}$ and 1$\times$10$^{14}$ cm$^{-2}$ (Fig. \ref{fig:SP}(c)).
$\Delta^{\rm SO}$ is the spin-orbit gap energy and is taken into account only in the calculation of the split-off holes.
$\Delta^{\rm SO}$ of diamond is 6 meV\cite{Win03}.
A secondary ion mass spectrometry measurement on a diamond substrate similar to the ones we used for fabricating the FETs indicated that the concentration of nitrogen, which acts as a donor, is 0.5 ppm and that of boron, which acts as an acceptor, is 5 ppb. Although these values may vary in the substrate and there may be defects ({\it e.g.} vacancies) that influence the value of $N_{\rm depl}$, we assumed $N_{\rm depl}$ = 0.5 ppm.
We used the following effective masses obtained from the Luttinger parameters\cite{Nak13} for the (111) diamond surface:
\begin{eqnarray}
 m^{\rm HH}_{z}/m_{0} = 1/(\gamma_{1} - 2\gamma_{3}) &=& 0.763,
 \label{eq:mzHH} \\
 m^{\rm LH}_{z}/m_{0} = 1/(\gamma_{1} + 2\gamma_{3}) &=& 0.248,
 \label{eq:mzLH} \\
 m^{\rm SO}_{z}/m_{0} = 1/\gamma_{1} &=& 0.375,
 \label{eq:mzSO} \\
 m^{\rm HH}_{//}/m_{0} = 1/(\gamma_{1} + \gamma_{3}) &=& 0.299,
 \label{eq:mcHH} \\
 m^{\rm LH}_{//}/m_{0} = 1/(\gamma_{1} - \gamma_{3}) &=& 0.503,
 \label{eq:mcLH} \\
 m^{\rm SO}_{//}/m_{0} = 1/\gamma_{1} &=& 0.375,
 \label{eq:mcSO}
\end{eqnarray}
where $m_{z}^{i}$ is the effective mass along the $z$ direction perpendicular to the diamond surface and $m_{//}^{i}$ is the effective mass parallel to the diamond surface.

The above self-consistent calculation was performed for a given hole density $p_\mathrm{2D}$. The gate voltage $V_\mathrm{GS}$ corresponding to the hole density is expressed as
\begin{eqnarray}
V_\mathrm{GS} = -\frac{e(p_\mathrm{2D} + n_\mathrm{depl} - n_\mathrm{imp}) t_\mathrm{hBN}}{\epsilon_\mathrm{hBN}} + \psi_\mathrm{s} + \phi_\mathrm{ms}
\end{eqnarray}
Here, $n_\mathrm{depl}$ ($n_\mathrm{depl} = N_\mathrm{depl}z_\mathrm{depl}$, where $z_\mathrm{depl}$ is the depletion-layer thickness) is the sheet density of the fixed charge in the depletion layer and $n_\mathrm{imp}$ is the sheet density of charged impurities at the diamond surface (see below). $t_\mathrm{hBN}$ and $\epsilon_\mathrm{hBN}$ are the thickness and dielectric constant of the h-BN gate dielectric. $\psi_\mathrm{s}$ ($\textless0$) is the surface potential (relative to deep inside the diamond). 
$\phi_\mathrm{ms}= \phi_\mathrm{m} - \phi_\mathrm{s}$ is the difference between the work function ($\phi_\mathrm{m}$) of the metal gate and that ($\phi_\mathrm{s}$) of hydrogen-terminated diamond. 
$\phi_\mathrm{s} \approx 0.3$ eV, because the electron affinity of hydrogen-terminated diamond is $-1.3$ eV\cite{Cui98} and the Fermi level in the bulk diamond with $N_\mathrm{D}$ (nitrogen) = 0.5 ppm and $N_\mathrm{A}$ =5 ppb is $-1.6$ eV below the conduction band minimum. 
$\phi_\mathrm{m}$ for titanium (Au/Ti gate) is 4.3 eV\cite{Mic77}. 
Therefore, $\phi_\mathrm{ms} \approx 4.0$ eV. 
For the given $p_\mathrm{2D}$, $n_\mathrm{depl}$ and $\psi_\mathrm{s}$ were evaluated by performing the Schr\"{o}dinger-Poisson calculation. $n_\mathrm{imp}$ was determined by comparing the experimental and calculated mobilities, as described below. 
An example of the deduced $V_\mathrm{GS}-p_\mathrm{2D}$ relationship is shown in Fig. \ref{fig:SP}(d). The $V_\mathrm{GS}-p_\mathrm{2D}$ curve is in good agreement with the experimental one if a slightly different value of $N_{D}$ (${\approx} 2$ ppm) is used; there may be such a level of spatial variation in $N_D$ in an HPHT (high pressure high temperature) IIa diamond substrate.

Next, we calculated the scattering rate for heavy, light, and split-off holes. 
The calculation took into account four scattering mechanisms: surface impurity scattering, background ionized impurity scattering, acoustic phonon scattering, and surface roughness scattering.
Optical phonon scattering was not considered because the optical phonon energy is as large as 165 meV and the occupation number of optical phonons is small at room temperature in diamond\cite{Per10}.

The scattering rate equations for the four different mechanisms are described below.

\vspace{\baselineskip}
\noindent
A. Surface charged impurity scattering

The carriers are scattered by the Coulomb potential arising from charged impurities on the surface.
The scattering rate due to charged impurities at a distance $d$ above the diamond surface is given by
\begin{eqnarray}
 \frac{1}{\tau _{\rm imp}^{i}} = n_{\rm imp}
\frac{m_{//}^{i}}{2\pi \hbar ^{3}(k_{F}^{i})^{3}}
\left(\frac{e^{2}}{2\epsilon _{0} \epsilon _{s}}\right)^{2}
\int_{0}^{2k_{F}^{i}}
[F^{i}(q)]^{2}
\frac{\exp (-2qd)}{[q+q_{TF}^{i}G^{i}(q)]^{2}}
\frac{q^{2}dq}{\sqrt{1-(q/2k_{F}^{i})^{2}}},
 \label{eq:si} \\
 F^{i}(q) = \int_{0}^{\infty}dz\left|\Psi_{1}^{i}(z)\right|^{2}\exp (-qz),
 \label{eq:FF1} \\
 G^{i}(q) = \int_{0}^{\infty}dz\int_{0}^{\infty}dz'[\Psi_{1}^{i}(z)]^{2}[\Psi_{1}^{i}(z')]^{2}\exp(-q|z-z'|).
 \label{eq:FormFactor}
\end{eqnarray}
Here, $n_{\rm imp}$ is the density of surface charged impurities, $\hbar$ is the reduced Planck constant, $q_{TF}^{i} = m_{//}^{i}e^{2}/(2\pi \epsilon _{0}\epsilon _{s}\hbar ^{2})$ is the Thomas-Fermi screening wave vector, $F^{i}(q)$ and $G^{i}(q)$ are form factors, $k_{F}^{i} = \sqrt{2\pi p^{i}}$ is the Fermi wave vector, and $d$ is the distance between carriers and surface charged impurities.
In order to calculate the form factors, we used the wave function of the first subband of HH, LH, and SO obtained from the self-consistent calculation of the Schr\"{o}dinger and Poisson equations.
When we used Fang-Howard wave functions with $b^{i} = [12me^{2}(N_{\rm depl}+11p^{i}/32)/(\hbar ^{2}\epsilon _{0}\epsilon _{s})]^{1/3}$, the results were qualitatively the same. 
In this case, $F^{i}(q)=[b^{i}/(b^{i}+q)]^{3}$, and $G^{i}(q)=1/8\{2[b^{i}/(b^{i}+q)]^{3}+3[b^{i}/(b^{i}+q)]^{2}+3[b^{i}/(b^{i}+q)]\}$.

\vspace{\baselineskip}
\noindent
B. Background ionized impurity scattering

Ionized impurities in the substrate (background ionized impurities) induce carrier scattering.
The scattering rate due to background ionized impurities is given by (Ref. \cite{Dav98})
\begin{eqnarray}
 \frac{1}{\tau _{\rm impbulk}^{i}} = n^{\rm (3D)}_{\rm imp}
\frac{m_{//}^{i}}{2\pi \hbar ^{3}(k_{F}^{i})^{3}}
\left(\frac{e^{2}}{2\epsilon _{0} \epsilon _{s}}\right)^{2}
\int_{0}^{2k_{F}^{i}} 
\frac{1}{(q+q_{TF}^{i})^{2}}
\frac{qdq}{\sqrt{1-(q/2k_{F}^{i})^{2}}}.
 \label{eq:bi}
\end{eqnarray}
Here, $n^{\rm (3D)}_{\rm imp}=N_{D}+N_{A}$ is the density of background ionized impurities; the donors and acceptors should be fully ionized near the surface because of band bending.

\vspace{\baselineskip}
\noindent
C. Acoustic phonon scattering

A phonon is a quantum of lattice vibration and causes carrier scattering.
There are two modes of lattice vibration: acoustic and optical. 
The room-temperature mobility of diamond is mainly affected by acoustic phonons\cite{Li18}.
The scattering rate due to acoustic phonons is given by
\begin{eqnarray}
 \frac{1}{\tau _{\rm ac}^{i}} = \frac{m_{//}^{i}k_{B}TD_{\rm ac}^{2}}{\rho u_{l}^{2}\hbar ^{3}}\int_{-\infty}^{\infty}\left|\Psi_{1}^{i}(z)\right|^{2}\left|\Psi_{1}^{i}(z)\right|^{2}dz.
 \label{eq:ac}
\end{eqnarray}
Here, $D_{\rm ac}$ is the acoustic deformation potential, $\rho$ is the crystal mass density, and $u_{l}$ is the velocity of longitudinal acoustic phonons.
The deformation potential of diamond was calculated to be 8 eV by Cardona {\it et al.} using the linear combination of muffin tin orbitals (LMTO) method.\cite{Car86} This value is consistent with the temperature dependence of the mobility of boron-doped diamond found in the study by Pernot {\it et al.}\cite{Per10}.
$\rho$ and $u_{l}$ of diamond are 3515 kgm$^{-3}$ and 17536 ms$^{-1}$, respectively\cite{Per10}.
For the Fang-Howard wave functions, $\int_{-\infty}^{\infty}\left|\Psi_{1}^{i}(z)\right|^{2}\left|\Psi_{1}^{i}(z)\right|^{2}dz=3b^{i}/16$.

\vspace{\baselineskip}
\noindent
D. Surface roughness scattering

Surface roughness induces disorder in the electric potential and leads to carrier scattering.
The scattering rate due to surface roughness is given by
\begin{eqnarray}
 \frac{1}{\tau _{\rm sr}^{i}} = 
\frac{\Delta ^{2}\Lambda^{2}e^{4}m_{//}^{i}}{(\epsilon _{0}\epsilon _{s})^{2}\hbar ^{3}}
(p_{\rm 2D}+n_{\rm depl})^{2}
\int_{0}^{1}
\frac{u^{4}\exp [-(k_{F}^{i})^{2}\Lambda^{2}u^{2}]}{[u+G^{i}(2k_{F}^{i}u)q_{TF}^{i}/(2k_{F}^{i})]^{2}\sqrt{1-u^{2}}}du.
 \label{eq:sr}
\end{eqnarray}
The surface roughness is characterized by the average roughness ($\Delta$) and correlation length ($\Lambda$).

\vspace{\baselineskip}
The total scattering rate is calculated using the Mathiessen rule,
\begin{eqnarray}
 \frac{1}{\tau^{i}} = 
\frac{1}{\tau _{\rm imp}^{i}} + \frac{1}{\tau _{\rm impbulk}^{i}} + \frac{1}{\tau _{\rm ac}^{i}} + \frac{1}{\tau _{\rm sr}^{i}},
 \label{eq:tauall}
\end{eqnarray}
and the mobility for $i$=HH, LH, and SO is obtained from $\mu^{i} = e\tau^{i}/m_{//}^{i}$.
We calculated the carrier density and mobility of the FET using the formula for the multi-carrier Hall effect, because our experimental results were obtained from low-magnetic-field Hall-effect measurements,
\begin{eqnarray}
 \mu = \frac{p^{\rm HH}(\mu^{\rm HH})^{2}+p^{\rm LH}(\mu^{\rm LH})^{2}+p^{\rm SO}(\mu^{\rm SO})^{2}}{p^{\rm HH}\mu^{\rm HH}+p^{\rm LH}\mu^{\rm LH}+p^{\rm SO}\mu^{\rm SO}},
 \label{eq:muHall} \\
 p = \frac{(p^{\rm HH}\mu^{\rm HH}+p^{\rm LH}\mu^{\rm LH}+p^{\rm SO}\mu^{\rm SO})^{2}}{p^{\rm HH}(\mu^{\rm HH})^{2}+p^{\rm LH}(\mu^{\rm LH})^{2}+p^{\rm SO}(\mu^{\rm SO})^{2}}.
 \label{eq:nHall}
\end{eqnarray}

\section{results and discussion}
\begin{figure}
 \includegraphics[width=16.4truecm]{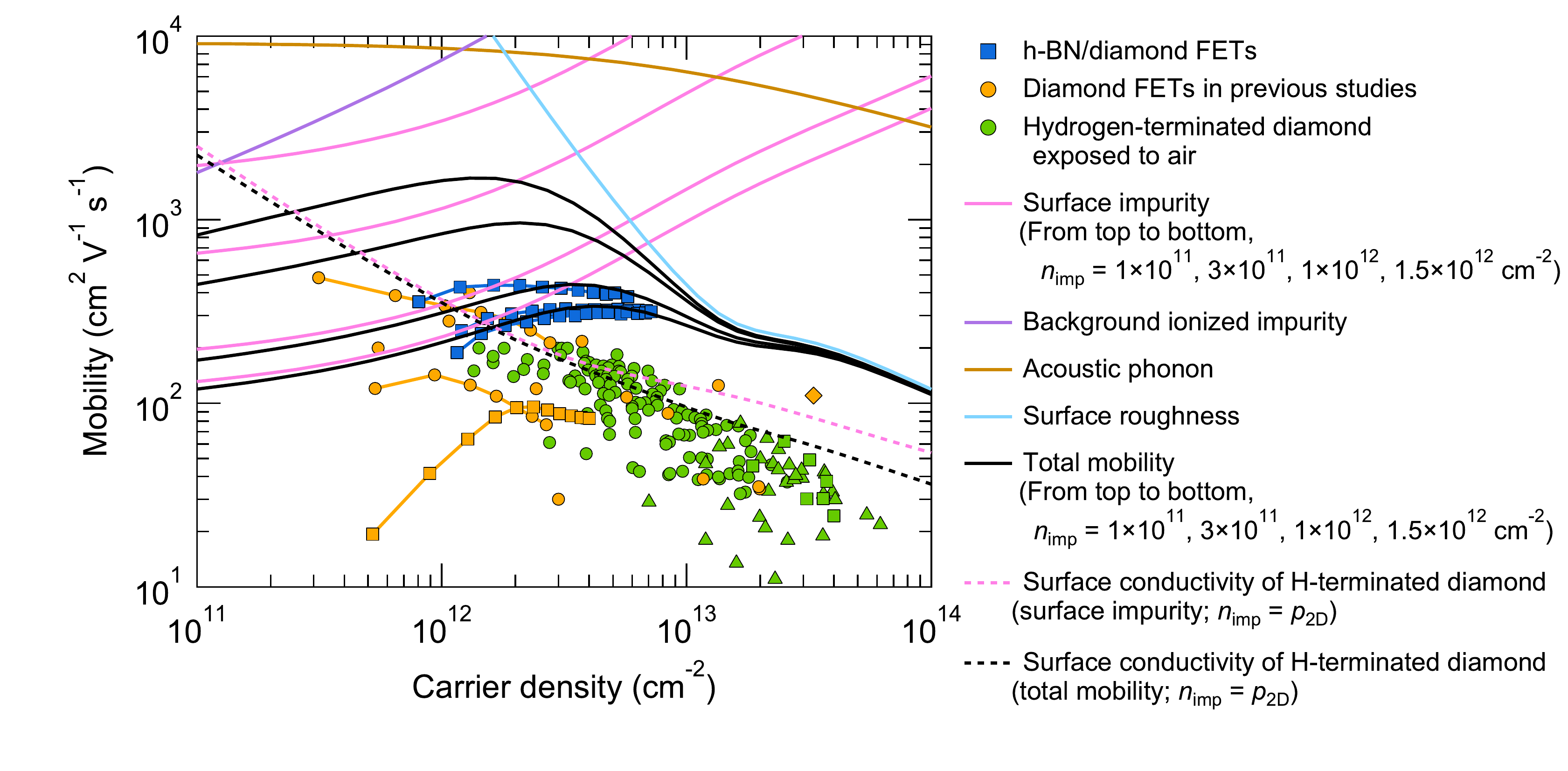}
 \caption{Carrier density dependence of mobility. Solid lines show calculated results for hydrogen-terminated diamond FETs. Dashed lines show calculated results for the surface conductivity of hydrogen-terminated diamond exposed to air. The figure also shows experimental results\cite{Sas18} for our h-BN/diamond FETs, other groups' diamond FETs and surface conductivities of hydrogen-terminated diamond exposed to air. Squares, circles, and triangles represent (111), (100), and (110) diamond surfaces, respectively. The rhombus represents polycrystalline diamond.}
 \label{fig:muall}
\end{figure}

Figure \ref{fig:muall} shows the carrier density dependence of mobility of our three diamond FETs with h-BN gate dielectric\cite{Sas18}. 
The figure also shows the mobilities of diamond FETs reported by other groups and the mobility of surface conductivity of hydrogen-terminated diamond surfaces exposed to air\cite{Sas18}.
The mobilities of our FETs exceeded 300 cm$^{2}$V$^{-1}$s$^{-1}$, and they were only weakly dependent on the carrier density. This contrasts with the mobilities of the surface conductivity induced by air exposure, which decreases monotonically with carrier density.

Here, let us identify which scattering mechanism limits the mobility of our FETs.
The carrier density dependence of the mobility limited by acoustic phonon scattering is basically determined by material-dependent parameters such as the deformation potential and phonon velocity.
As shown in Fig. \ref{fig:muall}, the calculated acoustic-phonon-limited mobility is more than one order of magnitude higher than the experimental one, and therefore, acoustic phonons are not the dominant scattering sources in our FETs.
The high acoustic-phonon-limited mobility is due to the high phonon velocity and the large crystal mass density in diamond.

The analytical formula for the surface roughness scattering contains device-dependent parameters, $\Delta$ and $\Lambda$, which characterize the magnitude of surface roughness.
Here, we will assume $\Delta$ = 0.3 nm and $\Lambda$ = 2 nm.
The validity of these values is described below.
The calculated mobility limited by the surface roughness scattering is also higher than the experimental one, and it is a strongly decreasing function of carrier density. Therefore, surface roughness scattering cannot explain the overall behavior of the mobility of our FETs, either.
Similarly, the background ionized impurity scattering cannot explain the mobility of our FETs because it leads to a mobility one order of magnitude higher than the experimental one.

\begin{figure}
 \includegraphics[width=8truecm]{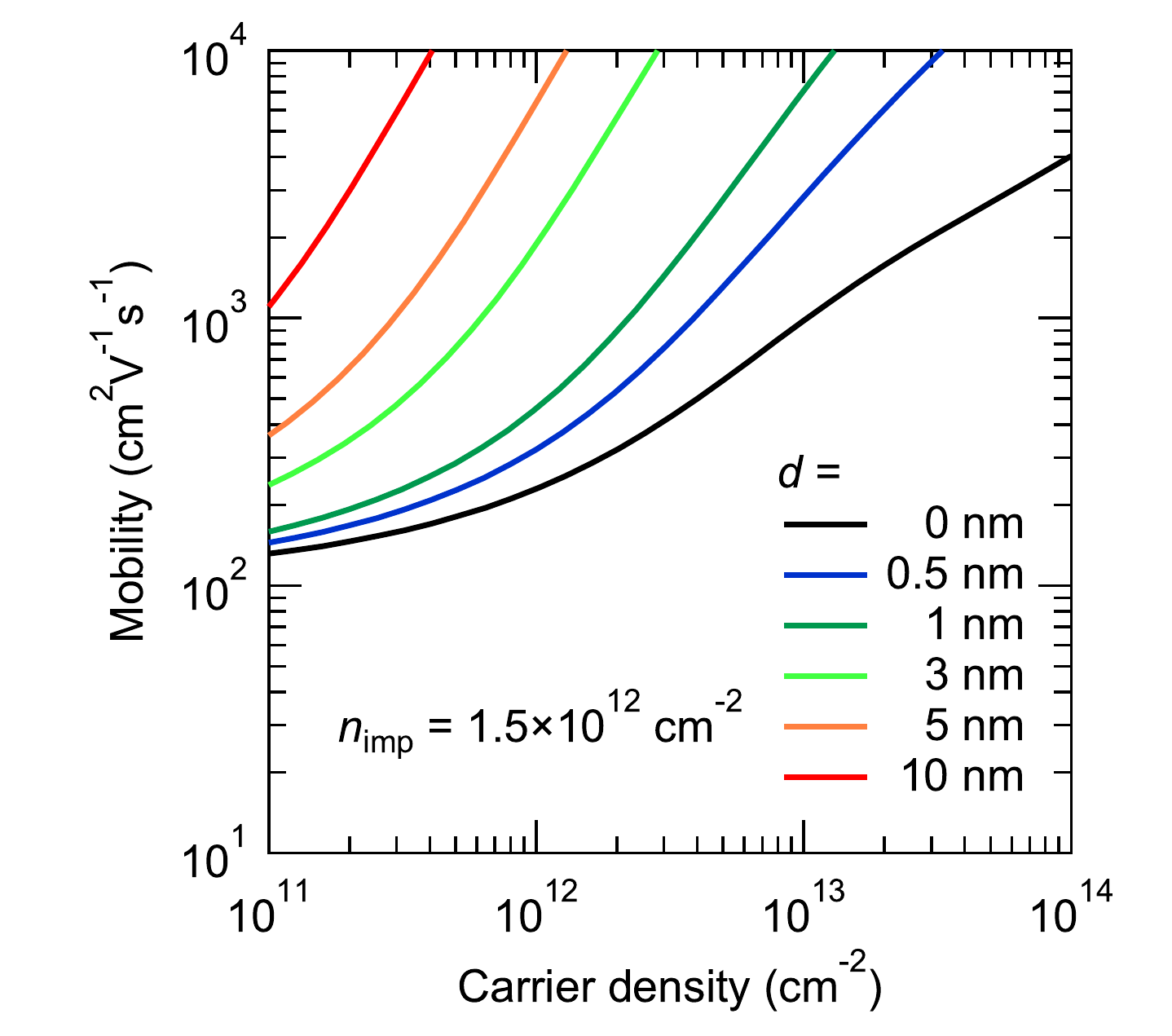}
 \caption{Calculated carrier density dependence of mobility limited by surface impurity scattering. The density of surface charged impurities $n_{\rm imp}$ is 1.5$\times$10$^{12}$ cm$^{-2}$. $d$ is the distance between the charged impurities and the two-dimensional hole gas.}
 \label{fig:mudDep}
\end{figure}

The surface impurity scattering with a constant impurity density leads to a slow increase in mobility with carrier density if the distance between the impurities and the two-dimensional hole gas is small (Fig. \ref{fig:mudDep}).
We calculated the mobility limited by surface charged impurities for different $n_{\rm imp}$ by assuming $d$ = 0.
The total mobility calculated with $n_{\rm imp}$ = (1.0-1.5)$\times$10$^{12}$ cm$^{-2}$ agrees reasonably well with the experimental mobility, as shown in Fig. \ref{fig:muall}.
The comparison between the experimental and calculated mobilities thus indicates that the surface impurity scattering is the dominant mechanism that limits the mobility of our FETs.

\begin{figure}
 \includegraphics[width=8truecm]{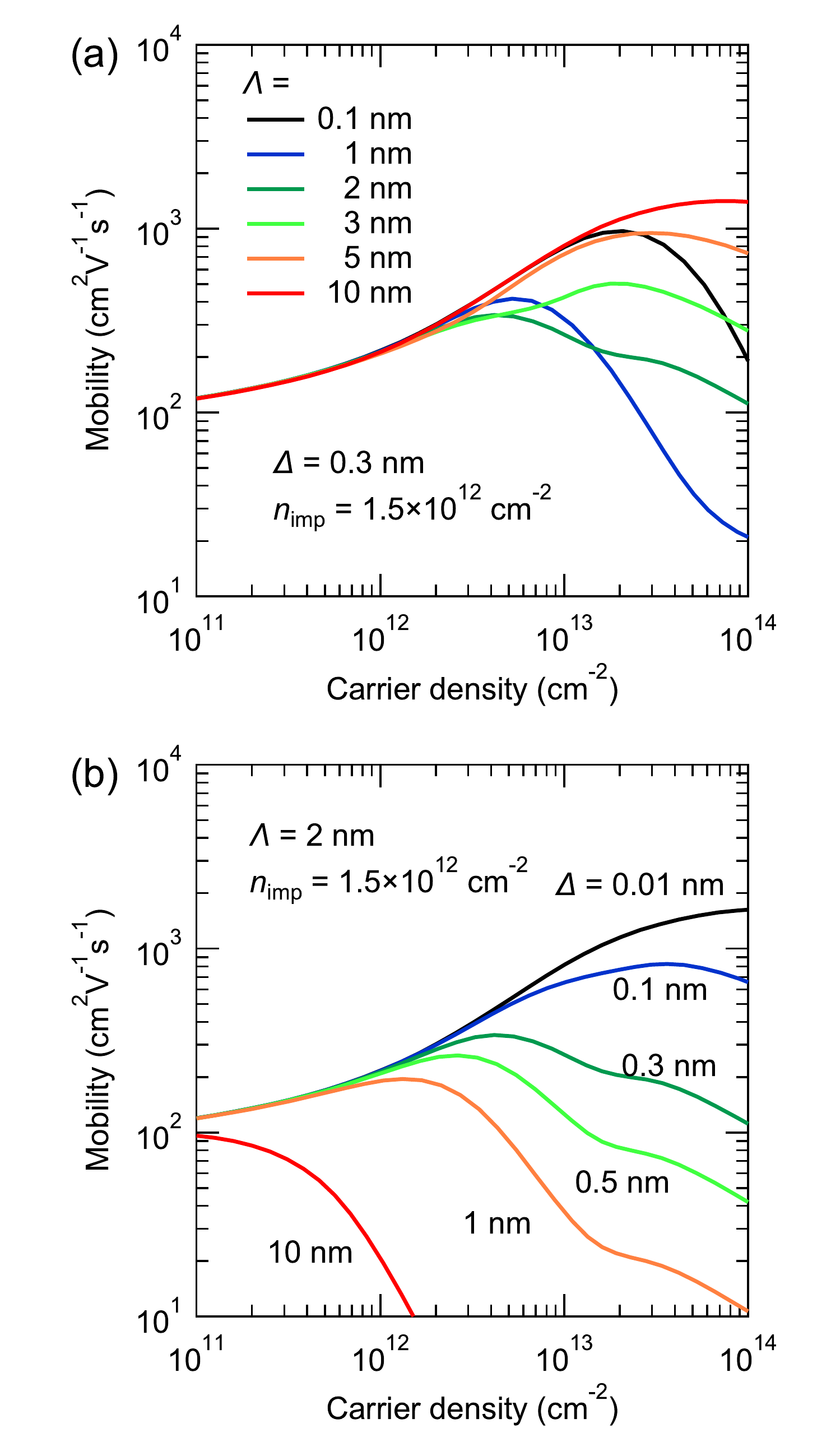}
 \caption{(a) Calculated carrier density dependence of mobility for different values of $\Lambda$. $\Delta$ is 0.3 nm, and $n_{\rm imp}$ is 1.5$\times$10$^{12}$ cm$^{-2}$. (b) Calculated carrier density dependence of mobility for different values of $\Delta$. $\Lambda$ is 2 nm, and $n_{\rm imp}$ is 1.5$\times$10$^{12}$ cm$^{-2}$.}
 \label{fig:muLdelDep}
\end{figure}

We assumed $\Delta$=0.3 nm and $\Lambda$=2 nm in the calculation of the surface roughness scattering rate. Taking different values of $\Lambda$ does not considerably influence the carrier density dependence of the mobility for carrier densities lower than 5$\times$10$^{12}$ cm$^{-2}$ (Fig. \ref{fig:muLdelDep}(a)).
The value of 2 nm for $\Lambda$ is the same as the one obtained by Li {\it et al.} This value is also comparable to those used for explaining the mobility in a Si MOSFET\cite{And82} and an AlGaN/GaN heterostructure\cite{Zan04}.
The value of $\Delta$, in contrast, significantly affects the carrier density dependence of the mobility, as shown in Fig. \ref{fig:muLdelDep}(b).
If $\Delta$ is larger than 1 nm, the mobility decreases rapidly with increasing carrier density. Such a rapid decrease is inconsistent with the experimental results.
The experimental mobility can be explained most reasonably with $\Delta\approx$0.3 nm.
This is within a typical range of surface roughness of polished diamond substrates.

\begin{figure}
 \includegraphics[width=8truecm]{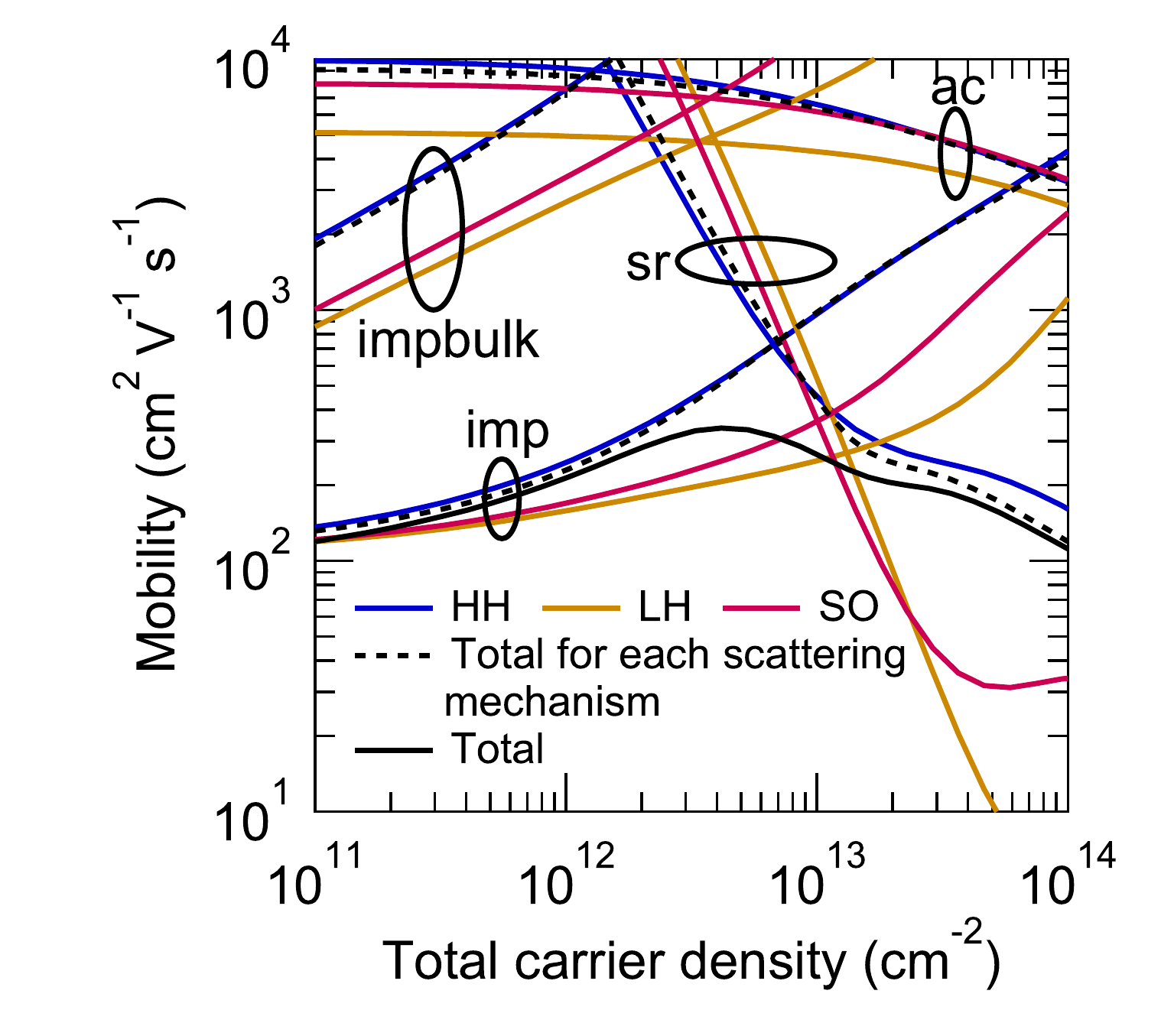}
 \caption{Calculated mobilities for heavy, light, and split-off holes as a function of the total hole density. The black solid line represents the total mobility. The black dashed lines represent the total mobility for each scattering mechanism. imp, impbulk, ac, and sr indicate surface impurity, background ionized impurity, acoustic phonon, and surface roughness scattering, respectively.}
 \label{fig:mu_HHLHSO}
\end{figure}

Figure \ref{fig:mu_HHLHSO} shows the mobilities of heavy, light, and split-off holes as a function of the total hole density. The calculated mobility of the heavy holes is almost the same as the total mobility for each scattering mechanism, indicating the dominant role of heavy holes in carrier transport. This is reasonable because the proportion of the density of heavy holes is the largest. Another feature shown in the figure is that among the mobilities limited by surface impurity scattering, the mobility of the heavy holes is higher than those of the light and split-off holes for a given $p_\mathrm{2D}$. This is mainly because the density and therefore $k_\mathrm{F}$ of the heavy hole subband are the largest. If we plot the impurity-limited mobilities of heavy, light, and split-off holes against each density ($p^\mathrm{HH}$, $p^\mathrm{LH}$, and $p^\mathrm{SO}$), the difference between them is small (35\% at maximum, which is even smaller than the difference in effective mass $m_{//}$).

Experimentally, the Hall mobility of our FETs was hardly dependent on temperature at $T\textgreater200$ K (Fig. 4 of Ref. 10), which is consistent with our model with the dominant contribution coming from impurity scattering. At lower temperature, however, the experimental mobility decreases with decreasing temperature (and shows an increase at $T\textless50$ K for high gate voltages). A similar decrease in mobility has been observed in Si MOSFETs\cite{Fan68}. As in the study of Si MOSFETs, we interpret the decrease in mobility to be a result of hole localization caused by potential fluctuations due to the surface charged impurities.

\begin{figure}
 \includegraphics[width=8truecm]{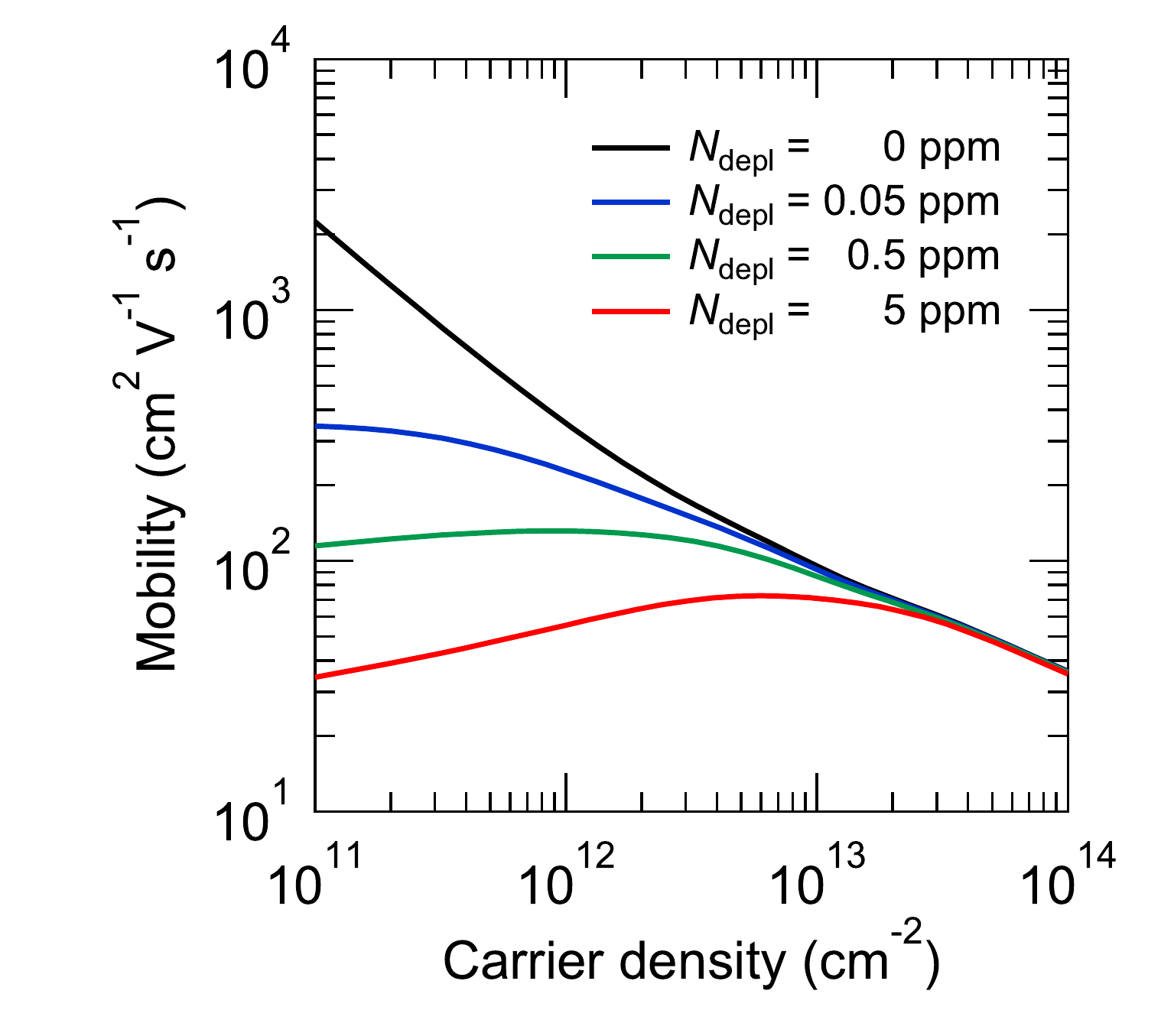}
 \caption{Calculated carrier density dependence of mobility for the surface conductivity of hydrogen-terminated diamond for different values of $N_\mathrm{D} = 0, 0.05, 0.5$ and $5$ ppm with $N_\mathrm{A}/N_\mathrm{D}$ kept to be 0.01. The calculation was performed assuming $n_{\rm imp} = p_{\rm 2D} + n_{\rm depl}$.}
 \label{fig:mu_SC}
\end{figure}

We also examined the mobility of the surface conductivity of the hydrogen-terminated surface exposed to air. The monotonic decrease in mobility with increasing carrier density (Fig. \ref{fig:muall}) can be explained almost quantitatively by surface impurity scattering with $n_{\rm imp} = p_{\rm 2D}$, as was reported by Li {\it et al.}\cite{Li18}. As an increase in carrier density means an increase in the density of charged impurities, the mobility monotonically decreases with carrier density.
In the carrier density range between 10$^{11}$ and 10$^{14}$ cm$^{-2}$, the surface impurity scattering leads to lower mobility than those limited by acoustic phonons and surface roughness; therefore, the surface impurity scattering is dominant.
We should note that our calculation used the effective masses for a (111) surface, although Fig. \ref{fig:muall} also shows experimental results for (100) surfaces. The difference in calculated mobility between the (100) and (111) surfaces is less than 40\% and is within the variation of the experimental mobilities.
We should also note that the contribution of $N_{\rm depl}$ is not considered above; more accurately, $n_{\rm imp}$ should equal $p_{\rm 2D} + n_{\rm depl}$.
We did not find the values of $n_{\rm depl}$ of the diamond samples in the literature from which the mobility of the surface conductivity were taken for plotting Fig. \ref{fig:muall}.
The above Schr\"{o}dinger-Poisson calculations show that $n_{\rm depl}=1.5\times10^{12}$ cm$^{-2}$ for $N_{\rm depl} = 8.8\times10^{16}$ cm$^{-3}$ (0.5 ppm) and $n_{\rm depl}=4.7\times10^{12}$ cm$^{-2}$ for $N_{\rm depl} = 8.8\times10^{17}$ cm$^{-3}$ (5 ppm).
If $n_\mathrm{depl}$ is taken into account, the calculated mobility decreases with increasing $n_\mathrm{depl}$ especially for low $p_\mathrm{2D}$ (Fig. \ref{fig:mu_SC}). The dependence of mobility on $n_\mathrm{depl}$ may partly explain why there is a wide distribution of mobilities of surface conductivity in the literature.

The above comparison between the experimental and calculated mobility indicates that the surface impurity scattering is the dominant mechanism that limits the mobility of our FETs.
This is consistent with our recent finding that the quantum and transport lifetimes estimated from Shubnikov-de Hass oscillations at low temperatures are nearly the same\cite{Sas19}.
The surface charged impurities may be adsorbed when the diamond surface is exposed to air before it is laminated by a flake of h-BN\cite{Sas18}.
Most of the heterostructures consisting of graphene and h-BN are also created by stacking the layers with their surfaces exposed to air. The interfaces of the layers can nevertheless be free from adsorbates due to unique self-cleansing effects.\cite{Kre14}
However, such self-cleansing effects seem to be ineffective for the interface between the hydrogen-terminated diamond and h-BN. 
To improve the mobility of the FETs, it would be important to reduce the density of adsorbates, for example, by a vacuum annealing\cite{Ina19}.
As shown in Fig. \ref{fig:muall}, our calculation suggests that decreasing the density of charged impurities to $\approx$1$\times$10$^{11}$ cm$^{-2}$ would lead to a mobility above 1000 cm$^{2}$V$^{-1}$s$^{-1}$ at room temperature.
Surface roughness scattering should also be reduced in the case of carrier densities higher than $\approx4\times10^{12}$ cm$^{-2}$.
For this purpose, it will be effective to use an atomically flat diamond surface prepared by chemical vapor deposition with a low methane concentration on a mesa structure.\cite{Tok08,Yam14}

\section{conclusions}
In conclusion, we calculated the carrier density dependence of the mobility of hydrogen-terminated diamond FETs in consideration of four scattering mechanisms: surface impurity scattering, background ionized impurity scattering, acoustic phonon scattering, and surface roughness scattering.
The calculated mobility agrees with the measured mobility of our diamond FETs with a h-BN gate dielectric if we assume a constant surface impurity density $n_{\rm imp} = (1.0-1.5)\times10^{12}$ cm$^{-2}$, average surface roughness $\Delta$ of 0.3 nm, and correlation length $\Lambda$ of 2 nm.
Decreasing the surface impurity density below $\approx$1$\times$10$^{11}$ cm$^{-2}$ will lead to a mobility exceeding 1000 cm$^{2}$V$^{-1}$s$^{-1}$. 
The mobility is significantly higher than that of $p$-type Si MOSFETs and will be useful for developing electronic devices that operate with low loss and high speed.

\begin{acknowledgments}

We thank T. Teraji and S. Koizumi for their helpful discussions. This study was supported by Grants-in-Aid for Scientific Research (Grants Nos. 25287093, 26630139, 19J12696 and 19H02605) and the ``Nanotechnology Platform Project'' of MEXT, Japan.

\end{acknowledgments}

\bibliography{diamondFET}

\newpage

\end{document}